\documentclass[10pt, conference, letterpaper]{IEEEtran}
\usepackage[letterpaper, left=0.65in, right=0.65in, bottom=1.05in, top=0.75in]{geometry}
\IEEEoverridecommandlockouts
\usepackage{cite}
\usepackage{amsmath,amssymb,amsfonts}
\usepackage{graphicx}
\usepackage[caption=false,font=footnotesize]{subfig}
\usepackage{textcomp}
\usepackage{xcolor}
\usepackage{tabularx}
\usepackage{multirow}
\usepackage{diagbox}
\usepackage[binary-units]{siunitx} 
\DeclareSIUnit{\bps}{bps}
\usepackage{amsthm}
\usepackage{mathrsfs} 
\begin{document}

\title{ Heterogeneous System Design for Cell-Free Massive MIMO in Wideband Communications }

\author{\IEEEauthorblockN{Wei Jiang\IEEEauthorrefmark{1} and Hans Dieter Schotten\IEEEauthorrefmark{2}}
\IEEEauthorblockA{\IEEEauthorrefmark{1}German Research Center for Artificial Intelligence (DFKI)\\
  }
\IEEEauthorblockA{\IEEEauthorrefmark{2}University of Kaiserslautern (RPTU)\\
 }
}
\maketitle

\begin{abstract}
Cell-free massive multi-input multi-output (CFmMIMO) offers uniform service quality through distributed access points (APs), yet unresolved issues remain. This paper proposes a heterogeneous system design that goes beyond the original CFmMIMO architecture by exploiting the synergy of a base station (BS) and distributed APs. Users are categorized as near users (NUs) and far users (FUs) depending on their proximity to the BS. The BS serves the NUs, while the APs cater to the FUs. Through activating only the closest AP of each FU, the use of downlink pilots is enabled, thereby enhancing performance.  This heterogeneous design outperforms other homogeneous massive MIMO configurations, demonstrating superior sum capacity while maintaining comparable user-experienced rates. Moreover, it lowers the costs associated with AP installations and reduces signaling overhead for the fronthaul network.
\end{abstract}

\section{Introduction}
Recently, cell-free massive multi-input multi-output (CFmMIMO) \cite{Ref_ngo2017cellfree} has attracted considerable attention from both academia and industry. It is capable of offering uniform quality of service (QoS) for all users, effectively addressing the under-served problem commonly encountered at the edges of conventional cellular networks. There are no cells and cell boundaries. Instead, a large number of distributed access points (APs) simultaneously serve a relatively smaller number of users. Alongside the \textit{all-participating} CFmMIMO strategies, Buzzi \textit{et al.} proposed a \textit{user-centric} approach for cell-free massive MIMO (UCmMIMO) \cite{Ref_buzzi2017cellfree, Ref_buzzi2020usercentric}. In this approach, each AP serves only a subset of users in close proximity, reducing the amount of fronthaul overhead while maintaining comparable performance.

Despite its high potential, there exist several issues yet unresolved. First, the cell-free architecture poses high implementation costs, as hundreds of wireless sites must be identified for AP installations, and large-scale fiber cables are required to interconnect these APs \cite{Ref_masoumi2020performance}.  Second, in contrast to voice-centric cellular networks like GSM in the 1990s, which prioritized uniform QoS for voice calls, modern and next-generation systems must provide differentiated QoS tailored to the specific demands of diverse applications \cite{Ref_jiang2021road}. Essentially, while the worst-case QoS is improved in CFmMIMO, the performance of some other users is compromised through averaging. Third, earlier studies on CFmMIMO have typically assumed flat fading channels, conducting algorithm design and performance analyses within a \textit{coherence interval}. This assumption holds only in narrow-band communications. Nowadays, however, most wireless systems operate in wideband, with signal bandwidths far exceeding the \textit{coherence bandwidth} \cite{Ref_jiang2021cellfree}. 

Responding to these issues, we propose a heterogeneous system design for CFmMIMO in wideband communications. It seamlessly integrates a base station (BS) and distributed APs in a cell-free system. Users are categorized into two groups: near users (NUs) and far users (FUs), depending on their proximity to the BS. The BS serves the NUs, while the APs cater to the FUs. Leveraging the frequency domain offered by orthogonal frequency-division multiplexing (OFDM), the use of downlink pilots is enabled by opportunistically activating the closest AP of each FU while deactivating other APs. Compared with three benchmark massive MIMO (mMIMO) setups, it outperforms in terms of both per-user spectral efficiency (SE) and sum capacity. Meanwhile, the implementation costs associated with AP installations are substantially lowered because the number of APs needed to be installed and connected is much less. The signaling overhead in the fronthaul network is also reduced since only a subset of APs participates in communications.

This paper is organized as follows: The subsequent section introduces the system model. Section III elaborates on the communication process and analyzes performance in Section IV. Section V presents the numerical results, and finally, Section VI draws the conclusions.

\section{System Model}
In a cell-free system \cite{Ref_ngo2017cellfree}, a large quantity of $M$ APs are distributed across an intended coverage area. These APs are connected to a central processing unit (CPU) through a fronthaul network, coordinating their communications with $K$ user equipment (UE). The sets of APs and users are represented by $\mathbb{M}= \{1,\ldots,M\}$ and $\mathbb{K}=\{1,\ldots,K\}$, respectively, where $M\gg K$. Time-division duplexing (TDD) is employed to separate downlink (DL) and uplink (UL) signals, operating under the assumption that UL channel responses mirror those of the downlink due to channel reciprocity. This arrangement is necessitated by the impractical overhead associated with inserting DL pilots into the massive number of service antennas. All APs send DL signals over the same time-frequency resource, while the users simultaneously transmit in the uplink at a different instant \cite{Ref_nayebi2017precoding}. 

\begin{figure}[!bpht]
\centering
\includegraphics[width=0.39\textwidth]{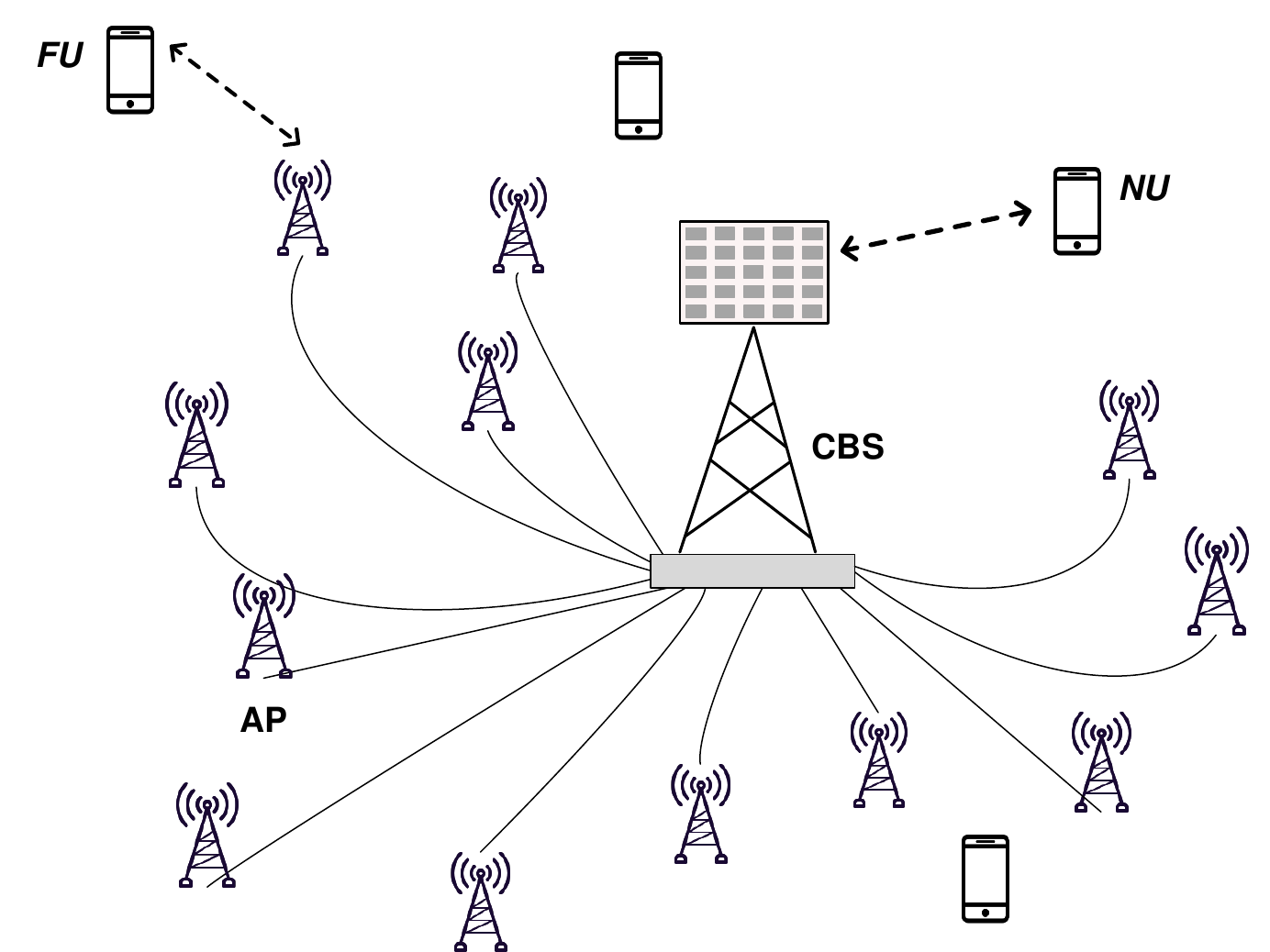}
\caption{The system model of HmMIMO consists of a CBS, APs, and UEs. } 
\label{Dia_SysModel}
\end{figure}

In contrast to the original cell-free architecture, we propose a heterogeneous design for CFmMIMO, referred to as HmMIMO. As illustrated in \figurename~\ref{Dia_SysModel}, a base station (BS) is centrally located, equipped with an array of $N_{b}$ antennas, which are collectively denoted as $\mathbb{M}_{bs}=\{1,\ldots,N_b\}$. Furthermore, $M-N_{b}$ single-antenna APs, represented by $\mathbb{M}_{ap}=\{N_b+1,\ldots,M\}$, are distributed randomly across the coverage area, similar to CFmMIMO. The BS also functions as the CPU for the APs. To distinguish it from conventional BSs, we henceforth refer to this combination of BS and CPU as a central BS (CBS). As evident, our design can lower implementation costs related to AP sites and fiber cables since the number of APs needed to be installed and connected is reduced.

Most previous CFmMIMO studies like \cite{Ref_ngo2017cellfree, Ref_interdonato2019downlink,  Ref_buzzi2017cellfree, Ref_zeng2021pilot} assume narrow-band communications under flat fading channels. Nevertheless, most modern wireless systems operate in wideband, with signal bandwidths far exceeding the coherence bandwidth. From a practical perspective, this paper considers \textit{frequency-selective fading} channels in wideband communications. The channel response between a typical antenna $m$ (either in the CBS or APs) and user $k$ is modeled as a linear time-varying filter in a baseband equivalent basis  \cite{Ref_tse2005fundamental}:  
\begin{align} \nonumber
\mathbf{g}_{mk}[t]&=\Bigl[ g_{mk,0}[t],\ldots,g_{mk,l}[t],\ldots, g_{mk,L_{mk}-1}[t]  \Bigr] ^T.
\end{align}
The filter length $L_{mk}$ should be no less than the multi-path delay spread $T_{d,mk}$ normalized by the sampling interval $T_s$, namely, $L_{mk} \geqslant  \lceil \frac{T_{d,mk}}{T_s} \rceil $.  The tap gain is given by $g_{mk,l}[t]=\sqrt{\beta_{mk}} h_{mk,l}[t]$, where $\beta_{mk}$ stands for large-scale fading, $h_{mk,l}[t] = \sum_i a_i(tT_s)e^{-2\pi j f_c \tau_i(tT_s)} \mathrm{sinc}\left[l - \frac{\tau_i(tT_s)}{T_s}\right]$, $f_c$ represents the carrier frequency, $a_i(tT_s)$ and  $\tau_i(tT_s)$ denote the attenuation and delay of the $i^{th}$ signal path, respectively, and $\mathrm{sinc}(x)\triangleq \frac{\sin(x)}{x}$ for $x\neq 0$. Large-scale fading experiences slow variations and is generally independent of frequency, making its acquisition and distribution simple. Therefore, $\beta_{mk}$, $\forall m\in\mathbb{M}$ and $k\in\mathbb{K}$, are assumed to be perfectly known \textit{a priori}. 

OFDM is an efficient technology to transmit wideband signals over frequency-selective channels, where the transmission is organized into blocks \cite{Ref_jiang2016ofdm}. Define $\tilde{x}_{m}[t,n]$ as the frequency-domain symbol conveyed on the $n^{th}$ subcarrier of the $t^{th}$ OFDM symbol at antenna $m$, resulting in a transmission block as $\tilde{\mathbf{x}}_m[t] = \left[ \tilde{x}_{m}[t,0],\ldots, \tilde{x}_{m}[t,N-1]  \right]^T$. Referring to \cite{Ref_jiang2021cellfree}, we know that the per-subcarrier DL signal model is given by
\begin{equation}
\label{Eqn_OFDMDL}
   \tilde{y}_{k}[t,n]= \sqrt{p_d} \sum_{m\in \mathbb{M}} \tilde{g}_{mk}[t, n]\tilde{x}_{m}[t,n]+\tilde{z}_{k}[t,n], \:\forall k\in\mathbb{K}, 
\end{equation}
where $\tilde{y}_{k}[t,n]$, $\tilde{g}_{mk}[t,n]$, and $\tilde{z}_{k}[t,n]$ represent the frequency-domain received symbol, channel response, and noise over the $n^{th}$ subcarrier of the $t^{th}$ OFDM symbol at user $k$.
Meanwhile, the per-subcarrier UL signal model is expressed as
\begin{equation} \label{eqn:ULKTranX}
   \tilde{r}_{m}[t,n]= \sqrt{p_u} \sum_{k\in \mathbb{K}}\tilde{g}_{mk}[t,n]\tilde{s}_{k}[t,n]+\tilde{z}_{m}[t,n], \:\forall m\in\mathbb{M},  
\end{equation} 
where $\tilde{r}_{m}[t,n]$, $\tilde{s}_{k}[t,n]$, and $\tilde{z}_{m}[t,n]$ are the frequency-domain received symbol at antenna $m$, transmitted symbol at user $k$, and noise, over the $n^{th}$ subcarrier of the $t^{th}$ OFDM symbol. $p_d$ and $p_u$ are per-antenna and per-user power constraints, respectively.

\section{The Communication Protocols}
The operation of HmMIMO lies in three core mechanisms:
\begin{itemize}
    \item The CBS labels each user as NU or FU based on factors like its distance to the CBS.
    \item In the uplink, all users simultaneously transmit over the same time-frequency resource. The signals of the NUs are detected by the CBS, treating the FUs' signals simply as interference. The APs process the signals of the FUs while disregarding the NUs' signals.
    \item The DL transmission resources are orthogonally divided into two parts: one for the NUs and another for the FUs. The CBS exclusively delivers the data for the NUs over their assigned resources. The data for the FUs are collaboratively sent by a subset of APs, consisting of the closest AP of each FU, while other APs are turned off. 
\end{itemize}
This section will elaborate on the communication protocols involved in user classification, resource allocation, channel estimation, uplink, and downlink data transmission. 

\subsubsection{User Classification} First, the CBS categorizes each user as a near or far user, according to a certain criterion like its distance to the CBS. A possible method, for instance, is to use large-scale fading as a measure of distance, and then form a set of NUs as $\mathbb{K}_N=\{k \mid \beta^0_k \geqslant \bar{\beta}_0\}$, where $\beta^0_k$ expresses the large-scale fading between the CBS and user $k$ and $\bar{\beta}_0$ denotes a pre-defined threshold.  The remaining users form a set of FUs $\mathbb{K}_F=\{k\mid \beta^0_k < \bar{\beta}_0\}$.

\subsubsection{Resource Allocation} 
\begin{figure}[!bpht]
\centering
\includegraphics[width=0.35\textwidth]{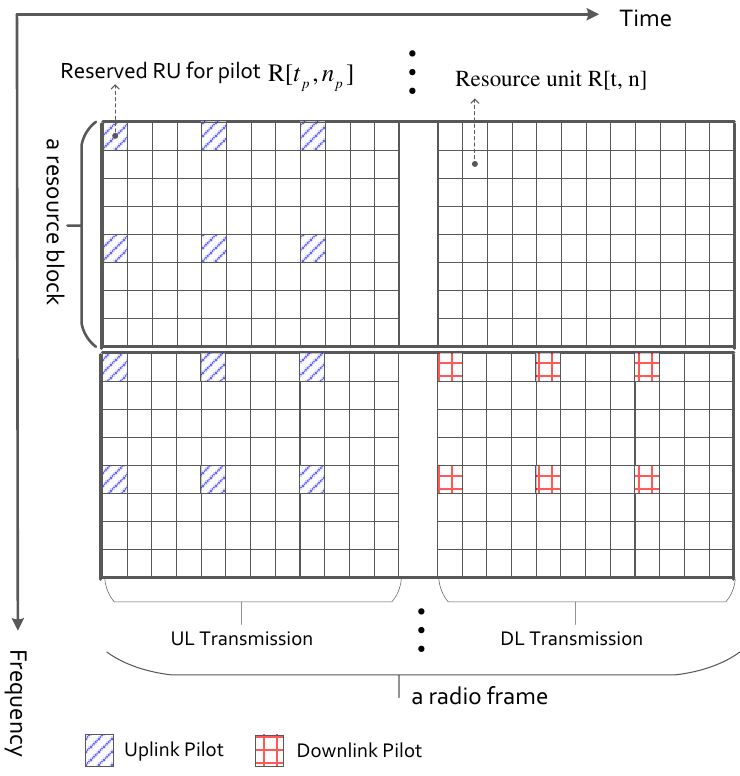}
\caption{Illustration of the OFDM time-frequency resource grid with two RB examples, each consisting of $8$ subcarriers spanning $24$ OFDM symbols. Users simultaneously transmit across all RBs with UL pilots, while DL resources for NUs, exemplified by the upper RB, lack pilot insertion.  Conversely, the activated $K$ FUs can insert DL pilots, as shown by the lower RB. } 
\label{Diagram_OFDMgrid}
\end{figure}

Assume a radio frame is comprised of $T$ OFDM symbols. Write $\mathscr{R} [ t,n ]$ to express a resource unit (RU) offered by the $n^{th}$ subcarrier of the $t^{th}$ OFDM symbol, where $n=0,\ldots,N-1$ and $t=0,\ldots,T-1$.  The granularity of resource allocation is specified as a resource block (RB), as shown in \figurename~\ref{Diagram_OFDMgrid}, encompassing $N_{rb}$ subcarriers throughout the entire duration of a radio frame. Consequently, there are $B=N/N_{rb}$ RBs, each consists of $T\times N_{rb}$ RUs. Write $\mathbb{B} \triangleq \left\{ \mathscr{R}[ t,n ] \mid 0\leqslant t < T/2, \: 0 \leqslant n<N \right\}$ to denote the resources for the UL transmission, assuming the equal UL/DL allocation is applied for simplicity. In the uplink, all users simultaneously transmit over the same time-frequency resources. However, the DL resources are orthogonally divided into two parts: $\mathbb{B}_N$ and $\mathbb{B}_F$, which are dedicated to the NUs and FUs, respectively. 

\subsubsection{Uplink Channel Estimation}
By properly inserting lattice-type pilot symbols, as illustrated in \figurename~\ref{Diagram_OFDMgrid}, the channel response of any RU can be directly estimated or interpolated by leveraging temporal and spectral correlation \cite{Ref_liu2014channel}. Under a \textit{block fading} model, the transmission of a radio frame is carried out within the coherent time, and the width of an RB is constrained to be smaller than the coherence bandwidth. Hence, the frequency-domain channel gain between AP $m$ and user $k$ across the $b^{th}$ RB, where $b\in \{0,1,\ldots, B-1\}$, can be expressed by $\tilde{g}_{mk}[b]$. Each RB needs to reserve $K$ RUs for UL pilots, denoted by $\{\mathscr{R} [t_p[k],n_p[k]] \mid  k\in \mathbb{K}\}$.  
According to \eqref{eqn:ULKTranX}, the received signal of antenna $m$ on $\mathscr{R} [t_p[k],n_p[k]] $ is 
\begin{equation} 
    \tilde{r}_{m}[t_p[k],n_p[k]] =\sqrt{p_u}\tilde{g}_{mk}[b] \tilde{I}+\tilde{z}_{m}[t_p[k],n_p[k]], 
\end{equation} 
where $\tilde{I}$ is a known frequency-domain symbol with $\mathbb{E}[\vert \tilde{I} \vert^2]=1$. 
Applying the minimum mean-square error (MMSE) estimation obtains an estimate of $\tilde{g}_{mk}[b]$ as
\begin{equation} \label{eqn:MMSE} 
    \hat{g}_{mk}[b] = \left(\frac{\sqrt{p_u}\beta_{mk}\tilde{I}^*}{p_u\beta_{mk}|\tilde{I}|^2 + \sigma_z^2}\right) \tilde{r}_{m}[t_p[k],n_p[k]].
\end{equation}
Its variance equals
\begin{align} \nonumber \label{eqn:MMSE_var}
&\mathbb{E}\left[\vert\hat{g}_{mk}[b]\vert^2\right]=\mathbb{E}\left[\hat{g}_{mk}[b]\hat{g}^*_{mk}[b]\right] \\ \nonumber &=\frac{p_u\beta_{mk}^2\mathbb{E}\left[\left \vert \sqrt{p_u} \tilde{g}_{mk}[b]\tilde{I}+\tilde{z}_{m}[t_p[k],n_p[k]] \right \vert^2\right]}{(p_u\beta_{mk} + \sigma_z^2)^2}\\
&=\frac{p_u\beta_{mk}^2}{p_u\beta_{mk} + \sigma_z^2}. 
\end{align}

As a result, AP $m$ gets the channel estimates with all users, represented by $\{\hat{g}_{mk}[b]\mid k\in \mathbb{K}\}$. The estimates follow complex normal distribution $ \mathcal{CN}\left(0,\alpha_{mk}\right)$, where $\alpha_{mk}=\frac{p_u\beta_{mk}^2}{p_u\beta_{mk} + \sigma_z^2}$. The estimation error $ \xi_{mk}[b] = \tilde{g}_{mk}[b]-\hat{g}_{mk}[b]$ follows $ \mathcal{CN}\left(0, \beta_{mk} -\alpha_{mk}\right)$. The channel between the CBS and user $k$ is expressed by $\tilde{\mathbf{g}}_k[b]=[\tilde{g}_{1k}[b],\ldots,\tilde{g}_{N_bk}[b]]^T$. Its estimate and estimation error are $\hat{\mathbf{g}}_{k}[b]\in \mathcal{CN}(\mathbf{0},\alpha^0_{k} \mathbf{I}_{N_{b}})$ and $\boldsymbol{\xi}_{k}[b] \in \mathcal{CN}(\mathbf{0},(\beta^0_{k}-\alpha^0_{k})\mathbf{I}_{N_{b}})$, given $\alpha^0_{k}=\frac{p_u(\beta^0_{k})^2}{p_u\beta^0_{k} + \sigma_z^2}$.

\subsubsection{Uplink Data Transmission}
All users simultaneously send their UL signals on $\mathbb{B}$. Refering to \eqref{eqn:ULKTranX}, we know that AP $m$ sees
\begin{equation} 
   \tilde{r}_{m}[t,n]= \sqrt{p_u} \sum_{k\in \mathbb{K}} \tilde{g}_{mk}[t,n]\tilde{s}_{k}[t,n]+\tilde{z}_{m}[t,n].  
\end{equation}
Aligning with \cite{Ref_ngo2017cellfree}, we apply matched filtering (MF), a.k.a. maximum-ratio combining, aiming to amplify the desired signal while disregarding inter-user interference (IUI). The APs only process their received signals to facilitate the recovery of the FUs' data. That is, AP $m$ multiplies $\tilde{r}_{m}[t,n]$ with the conjugate of its locally obtained channel estimates with the FUs, and then delivers $\{\hat{g}_{mk}^*[b]\tilde{r}_m[t,n] \mid k\in \mathbb{K}_F \}$ to the CBS. To detect the symbol from FU $k$, a soft estimate is formed by combining the pre-processed signals from all APs, i.e.,  
\begin{align} \nonumber \label{QN_ULsoftestimate} 
   \hat{s}_k[t,n] &= \sum_{m\in \mathbb{M}_{ap}} \hat{g}_{mk}^*[b]\tilde{r}_m[t,n] \\ \nonumber
   &= \sqrt{p_u} \sum_{m\in \mathbb{M}_{ap}} \hat{g}_{mk}^*[b]  \sum_{k'\in \mathbb{K}} \tilde{g}_{mk'}[t,n]\tilde{s}_{k'}[t,n]\\ 
   &+\sum_{m\in \mathbb{M}_{ap}} \hat{g}_{mk}^*[b] \tilde{z}_{m}[t,n].  
\end{align}
Meanwhile, the CBS observes
\begin{equation} 
   \tilde{\mathbf{r}}_0[t,n]= \sqrt{p_u} \sum_{k\in \mathbb{K}} \tilde{\mathbf{g}}_{k}[t,n]\tilde{s}_{k}[t,n]+\tilde{\mathbf{z}}_0[t,n],  
\end{equation}
where the noise vector $\tilde{\mathbf{z}}_0\in \mathcal{CN}(\mathbf{0},\sigma^2_z\mathbf{I}_{N_b})$. The CBS only detects the signals of the NUs, treating the FUs' signals as interference. 
For $k\in\mathbb{K}_N$, the CBS builds a soft estimate of
\begin{align}  \label{GS_uplinksoftestimate}
    &\hat{s}_k[t,n] = \hat{\mathbf{g}}_k^H[t,n] \tilde{\mathbf{r}}_0[t,n]\\ \nonumber
    &=  \sqrt{p_u} \hat{\mathbf{g}}_k^H[t,n] \sum_{k'\in \mathbb{K}}  \tilde{\mathbf{g}}_{k'}[t,n]\tilde{s}_{k'}[t,n]+\hat{\mathbf{g}}_k^H[t,n]\tilde{\mathbf{z}}_0[t,n].
\end{align}

\subsubsection{Downlink Data Transmission and Channel Estimation}
From a user's perspective, nearby APs offer strong signal strength, while those from distant APs are much weaker. Our proposal involves selecting a cluster of nearby APs for the FUs while deactivating the distant ones.
One possible approach is that each FU determines its closest AP with the largest large-scale fading. A set of $K$ near APs denoted by $\mathbb{M}_F$ is obtained. This strategy degrades (virtually) high-dimensional $M\times K$ massive MIMO to low-dimensional $K\times K$ MIMO since $M\gg K$ in massive MIMO systems. Thus, the use of DL pilots requires $K$ RUs, substantially lowering the overhead, compared to $M$ DL pilots in CFmMIMO. This allows the FU $k$ to acquire channel estimates $\{\hat{\mathfrak{g}}_{mk}[b] \mid m\in \mathbb{M}_{F}\}$, which follow $ \mathcal{CN}\left(0,\psi_{mk}\right) $ with $\psi_{mk}= \frac{p_d\beta_{mk}^2}{p_d\beta_{mk} + \sigma_z^2}$. 

The selected APs $m\in \mathbb{M}_F$ collaboratively transmit the symbols intended for the FUs, represented by $\{d_k[t,n] \mid k\in \mathbb{K}_F \}$, on $\mathbb{B}_F$. To spatially multiplex these symbols,  conjugate beamforming (CBF) is generally applied. The transmitted signal for AP $m\in \mathbb{M}_F$ on $\mathscr{R} [t,n] \in \mathbb{B}_F$ is given by
   \begin{equation} \label{eqn:DLMU_TxSyml}
    \tilde{x}_{m}[t,n] =  \sum_{k\in \mathbb{K}_F} \sqrt{\eta_{mk}} \hat{g}_{mk}^*[b] d_{k}[t,n],
\end{equation} where $0\leqslant \eta_{mk}\leqslant 1$ denotes the power-control coefficient, satisfying $\sum_{k\in \mathbb{K}_F} \eta_{mk}\leqslant 1$.  
Substituting \eqref{eqn:DLMU_TxSyml} into \eqref{Eqn_OFDMDL} yields the observation of $k\in \mathbb{K}_F $ as
\begin{align}  \label{EQN_dlcbfRxsignal}
   &\tilde{y}_{k}[t,n]=\\ \nonumber
   &\sqrt{p_d}\sum_{m\in \mathbb{M}_F} \tilde{g}_{mk}[b]  \sum_{k'\in \mathbb{K}_F} \sqrt{\eta_{mk'}} \hat{g}_{mk'}^*[b] d_{k'}[t,n] +\tilde{z}_{k}[t,n].
\end{align}

On $\mathbb{B}_N$, the CBS transmits the symbols $\{d_k[t,n] \mid k\in \mathbb{K}_N \}$ intended for the NUs. Using CBF, these symbols are spatially multiplexed as $\sum_{k\in \mathbb{K}_N} \mathbf{E}_k  \hat{\mathbf{g}}_{k}^*[t,n] d_k[t,n] $, where $\mathbf{E}_k$ is a $N_b\times N_b$ diagonal matrix with the $m^{th}$ diagonal element being $\sqrt{\eta_{mk}}$. Consequently, the NU $k$ sees
\begin{align} \label{GS_DL_RxSig}
    \tilde{y}_{k}[t,n]= \sqrt{p_d} \tilde{\mathbf{g}}_{k}^T[t,n] \sum_{k'\in \mathbb{K}_N} \mathbf{E}_{k'} \hat{\mathbf{g}}_{k'}^*[t,n] d_{k'}[t,n] + \tilde{z}_{k}[t,n].
\end{align}

\section{Performance Analysis}
This section analyzes the performance of HmMIMO in terms of per-user SE and sum capacity. For simplicity, the time and frequency indices of signals, i.e., $[t,n]$ and $[b]$, are omitted in the subsequent analysis.

\subsection{Uplink Performance}
Different availability levels of channel information correspond to different processing methods: coherent or non-coherent detection. The CBS knows $\hat{\mathbf{g}}_k$, $k\in \mathbb{K}$ by estimating the UL pilots. Hence, it is able to coherently detect the received signals to recover the NUs' information symbols. The achievable SE for $k\in \mathbb{K}_N$ is lower bounded by $R_{nu,k}^{ul}= \log(1+\gamma_{nu,k}^{ul})$, where the effective signal-to-interference-plus-noise ratio (SINR) equals
\begin{equation} \label{GS_UL_NU_SINR}
    \gamma_{nu,k}^{ul} =  \frac{   N_b  \alpha_k^{0}   }{   \sum_{k'\in \mathbb{K}}  \beta_{k'}^{0}    - \alpha_k^{0}   +  \frac{\sigma^2_z}{p_u}        }.
\end{equation}
\begin{IEEEproof}
The soft estimate in \eqref{GS_uplinksoftestimate} is decomposed to
\begin{align}  \label{EQN_UL_CBS_softesti} \nonumber
    \hat{s}_k & = \underbrace{ \sqrt{p_u} \| \hat{\mathbf{g}}_k  \|^2 \tilde{s}_{k} }_{\mathcal{S}_0:\:Desired\:signal} +\underbrace{ \sqrt{p_u} \hat{\mathbf{g}}_k^H  \boldsymbol{\xi}_{k}\tilde{s}_{k} }_{\mathcal{I}_1:\:Channel\:estimation\:error}\\
    &+ \underbrace{\sqrt{p_u} \hat{\mathbf{g}}_k^H \sum_{k'\neq k, k'\in \mathbb{K}}  \tilde{\mathbf{g}}_{k'}\tilde{s}_{k'} }_{\mathcal{I}_2:\:Inter-user\:interference} +\underbrace{\hat{\mathbf{g}}_k^H\tilde{\mathbf{z}}_0 }_{\mathcal{I}_3:\:Noise},
\end{align}
where $\mathcal{S}_0$, $\mathcal{I}_1$, $\mathcal{I}_2$, and $\mathcal{I}_3$ are mutually uncorrelated. According to \cite{Ref_hassibi2003howmuch}, the worst-case noise for mutual information is Gaussian additive noise with the variance equalling to the variance of $\mathcal{I}_1+\mathcal{I}_2+\mathcal{I}_3$. 
Thus, the effective SINR is calculated by
\begin{align}  \label{cfmmimo:formularSNR}
    \gamma  =  \frac{\mathbb{E}\left[|\mathcal{S}_0|^2\right]}{\mathbb{E}\left[|\mathcal{I}_1|^2\right]+\mathbb{E}\left[|\mathcal{I}_2|^2\right]+\mathbb{E}\left[|\mathcal{I}_3|^2\right]}
\end{align}
with 
\begin{align}  \label{APPEQ1}
    \mathbb{E}\left[|\mathcal{S}_0|^2\right] & = p_u \left( N_b \alpha_k^{0} \right)^2\\ \label{APPEQ2}
    \mathbb{E}\left[|\mathcal{I}_1|^2\right] & = p_u N_b \alpha_k^{0} (\beta_k^{0}-\alpha_k^{0})\\ \label{APPEQ3}
    \mathbb{E}\left[|\mathcal{I}_2|^2\right] & = p_u N_b \alpha_k^{0} \sum_{k'\neq k, k'\in \mathbb{K}}  \beta_{k'}^{0} \\  \label{APPEQ4}
    \mathbb{E}\left[|\mathcal{I}_3|^2\right] & = \sigma_z^2N_b \alpha_k^{0}.
\end{align}
Substituting these terms into \eqref{cfmmimo:formularSNR}, yields \eqref{GS_UL_NU_SINR}.
\end{IEEEproof}

Similar to the CPU in CFmMIMO, the CBS only possesses the full channel knowledge when each AP reports its local estimates. However, this method incurs significant signaling overhead. It is logical to presume that the CBS solely maintains channel statistics for the FUs, i.e.,  $ \mathbb{E} \left[ \left |  \hat{g}_{mk} \right | ^2\right]=\alpha_{mk}$, $k\in \mathbb{K}_F$. Consequently, detecting the FUs' signals coherently at the CBS is not possible. Transform \eqref{QN_ULsoftestimate} into
\begin{align}  \nonumber \label{GS_ULsoftestAP_channelUncertainty}
    \hat{s}_k  =&\underbrace{\sqrt{p_u} \sum_{m\in \mathbb{M}_{ap}} \mathbb{E}\left[| \hat{g}_{mk}|^2\right] \tilde{s}_k}_{\mathcal{S}_0} +\underbrace{\sqrt{p_u} \sum_{m\in \mathbb{M}_{ap}} \hat{g}_{mk}^*   \xi_{mk} \tilde{s}_k}_{\mathcal{I}_1}  \\  \nonumber
     &+ \underbrace{\sqrt{p_u} \sum_{m\in \mathbb{M}_{ap}} \hat{g}_{mk}^* \sum_{k'\neq k, k'\in \mathbb{K}}  \tilde{g}_{mk'} \tilde{s}_{k'} }_{\mathcal{I}_2}+   \underbrace{\sum_{m\in \mathbb{M}_{ap}} \hat{g}_{mk}^* \tilde{z}_m}_{\mathcal{I}_3}\\ & +\underbrace{\sqrt{p_u} \sum_{m\in \mathbb{M}_{ap}} \left( | \hat{g}_{mk}|^2 - \mathbb{E}\left[| \hat{g}_{mk}|^2\right] \right) \tilde{s}_k}_{\mathcal{I}_4:\:Channel\:uncertainty\:error}, 
\end{align} where an additional item $\mathcal{I}_4$ due to \textit{channel uncertainty} is imposed. 
The achievable SE for $k\in \mathbb{K}_F$ is lower bounded by $R_{fu,k}^{ul}= \log(1+\gamma_{fu,k}^{ul})$ with the effective SINR of 
\begin{equation} \label{GS_SINR_UL_AP}
    \gamma_{fu,k}^{ul} =  \frac{ \left( \sum_{m\in \mathbb{M}_{ap}} \alpha_{mk}  \right)^2}
    {\sum_{m\in \mathbb{M}_{ap}} \alpha_{mk} \sum_{k'\in \mathbb{K}}    \beta_{mk'}  +  \frac{\sigma^2_z}{p_u} \sum_{m\in \mathbb{M}_{ap}} \alpha_{mk}   }.
\end{equation} 
\begin{IEEEproof}
Likewise, in this case, we obtain
\begin{align} 
    \mathbb{E}\left[|\mathcal{S}_0|^2\right] & =  p_u  \left( \sum_{m\in \mathbb{M}_{ap}} \alpha_{mk}  \right)^2  \\ 
    \mathbb{E}\left[|\mathcal{I}_1|^2\right] & =  p_u   \sum_{m\in \mathbb{M}_{ap}} \alpha_{mk} (\beta_{mk}-\alpha_{mk})  \\ 
    \mathbb{E}\left[|\mathcal{I}_2|^2\right] & = p_u \sum_{m\in \mathbb{M}_{ap}} \alpha_{mk} \sum_{k'\neq k, k'\in \mathbb{K}}  \beta_{mk'}\\
    \mathbb{E}\left[|\mathcal{I}_3|^2\right] &= \sigma^2_z \sum_{m\in \mathbb{M}_{ap}} \alpha_{mk}\\
    \mathbb{E}\left[|\mathcal{I}_4|^2\right]  &= p_u \sum_{m\in \mathbb{M}_{ap}} \alpha_{mk}^2.
\end{align}
Thus, we get \eqref{GS_SINR_UL_AP}.
\end{IEEEproof}
The sum capacity of the HmMIMO system in the uplink is calculated by $ C_{ul}=\mathbb{B} \left( \sum_{k\in \mathbb{K}_N} R_{nu,k}^{ul}+\sum_{k\in \mathbb{K}_F} R_{fu,k}^{ul} \right) $.

\begin{figure*}[!tbph]
\centerline{
\subfloat[]{
\includegraphics[width=0.24\textwidth]{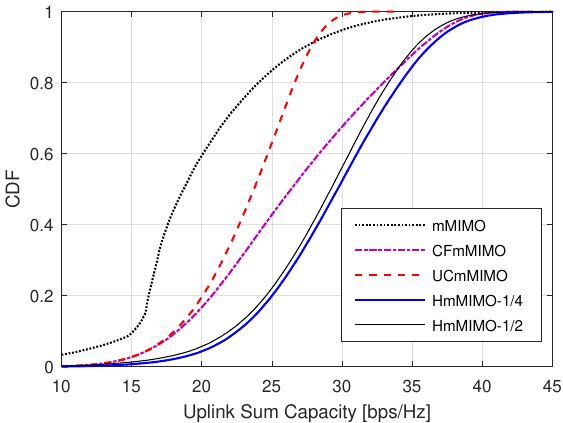}
\label{fig:result1} 
}
\hspace{-2mm}
\subfloat[]{
\includegraphics[width=0.24\textwidth]{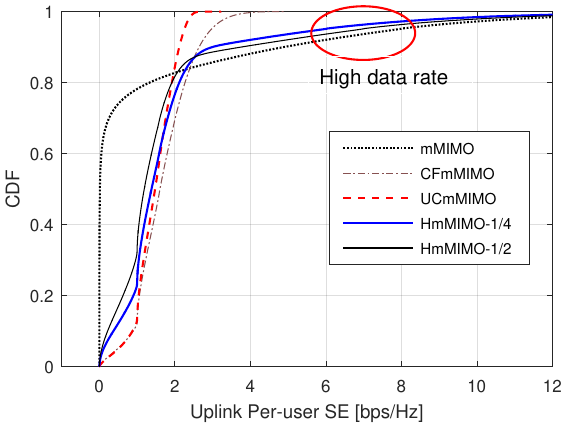}
\label{fig:result2} 
}
\hspace{-2mm}
\subfloat[]{
\includegraphics[width=0.24\textwidth]{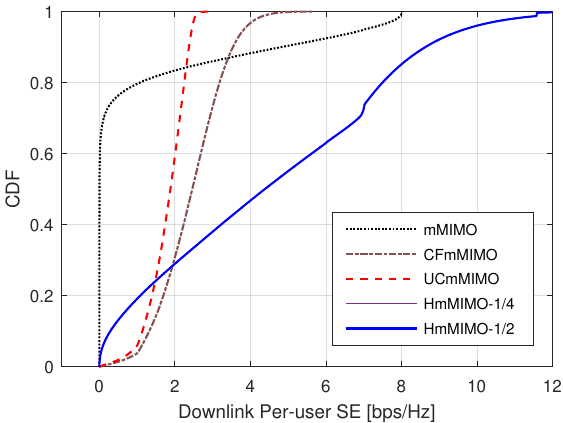}
\label{fig:result3}
}
\hspace{-2mm}
\subfloat[]{
\includegraphics[width=0.24\textwidth]{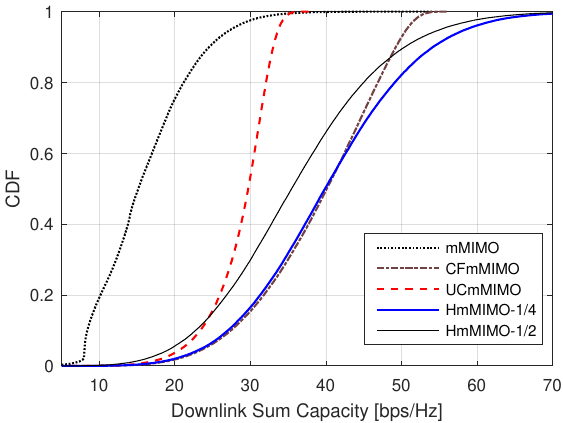}
\label{diagram:result2}
}
}
\hspace{15mm}
 \caption{Performance comparison of HmMIMO, CFmMIMO, UCmMIMO, and mMIMO, including (a) the CDF curves of UL sum capacity; (b) the CDF curves of UL per-user SE; (c) the CDF curves of DL per-user SE; and (d) the CDF curves of DL sum capacity;.    }
 \label{diagram:result}
\end{figure*}

\subsection{Downlink Performance}
The NUs know $\alpha_{k}^0$ rather than $\hat{\mathbf{g}}_{k}$ due to the lack of DL pilots, if  $N_b$ is large. Hence, $k\in\mathbb{K}_N$ cannot detect its received signals coherently. Further decompose \eqref{GS_DL_RxSig} into  
\begin{align} \nonumber
    \tilde{y}_{k} &= \sqrt{p_d }  \mathbb{E}\left[ \|\mathbf{E}_k\hat{\mathbf{g}}_{k} \|^2\right] d_{k} {+} \sqrt{p_d }  \left( \| \mathbf{E}_k \hat{\mathbf{g}}_{k} \|^2 {-} \mathbb{E}\left[ \| \mathbf{E}_k\hat{\mathbf{g}}_{k} \|^2\right] \right) d_{k}\\  \nonumber
    &+ \sqrt{p_d} \tilde{\mathbf{g}}_{k}^T \sum_{k'\neq k, k'\in \mathbb{K}_N} \mathbf{E}_k \hat{\mathbf{g}}_{k'}^* d_{k'}\\
    &+ \sqrt{p_d} \boldsymbol{\xi}_{k}^T \sum_{k'\in \mathbb{K}_N} \mathbf{E}_k \hat{\mathbf{g}}_{k'}^* d_{k'} + \tilde{z}_{k}.
\end{align}
Using similar manipulations as the derivation of uplink SE, we obtain the effective SINR of $k\in\mathbb{K}_N$ as 
\begin{equation} 
    \gamma_{nu,k}^{dl} =  \frac{  \left(\sum_{m=1}^{N_{b}} \sqrt{\eta_{mk}} \alpha_k^0 \right)^2   }
    {  \sum_{m=1}^{N_{b}} \beta_{k}^0 \sum_{k'\in \mathbb{K}_N} \eta_{mk'}  \alpha_{k'}^0 + \sigma^2_z/p_d    }.
\end{equation}
 
As the FUs conduct coherent detection with the aid of channel estimates, \eqref{EQN_dlcbfRxsignal} is rewritten to  
\begin{align}
\label{GS_downlink_Rx_Signal_for_FU} \nonumber
    \tilde{y}_k  &=  \sqrt{p_d}\sum_{m \in \mathbb{M}_F} \sqrt{\eta_{mk}} |\hat{g}_{mk} |^2 d_k \\  
    &+ \sqrt{p_d}\sum_{m \in \mathbb{M}_F} \sqrt{\eta_{mk}} \xi_{mk}  \hat{g}_{mk}^* d_k   \\ \nonumber 
     &+  \sqrt{p_d}\sum_{m \in \mathbb{M}_F} \tilde{g}_{mk} \sum_{k'\neq k, k'\in \mathbb{K}_F} \sqrt{\eta_{mk'}} \hat{g}_{mk'}^* d_{k'}   + \tilde{z}_k.
\end{align} 
Accordingly, we obtain the effective SINR for $k\in \mathbb{K}_F$ as 
\begin{align} \label{GS_SINR_DL_FU}
    &\gamma_{fu,k}^{dl} =\\ \nonumber
    &\frac{ \left( \sum_{m \in \mathbb{M}_F} \sqrt{\eta_{mk}} \alpha_{mk}\right)^2    }
    { \sum_{m \in \mathbb{M}_F} \beta_{mk} \sum_{k'\in \mathbb{K}_F } \eta_{mk'} \alpha_{mk'}   -\sum_{m \in \mathbb{M}_F}\eta_{mk} \alpha_{mk}^2 + \frac{\sigma^2_z}{p_d }   }.
\end{align}
The DL sum capacity of HmMIMO is computed by 
\begin{equation}
    C_{dl}=\mathbb{B}_N\sum_{k\in \mathbb{K}_N} \log(1+ \gamma_{nu,k}^{dl})+\mathbb{B}_F\sum_{k\in \mathbb{K}_F}\log(1+ \gamma_{fu,k}^{dl}).
\end{equation}

\section{Numerical Results}

The performance of HmMIMO is numerically evaluated in terms of per-user SE and sum capacity. Prior methods were designed for a coherence interval, neglecting the frequency selectivity in wideband communications.  To facilitate a fair comparison, \textit{mMIMO} with collocated antenna arrays \cite{Ref_yang2013performance}, \textit{CFmMIMO} \cite{Ref_ngo2017cellfree}, and \textit{UCmMIMO} \cite{Ref_buzzi2017cellfree} are extended to each OFDM subcarrier, serving as benchmarks. In our simulations, a representative scenario is established where $M=256$ antennas serve $K=16$ users. In CFmMIMO and UCmMIMO, all APs and users are randomly distributed across a circular area with a $1\mathrm{km}$ radius. In mMIMO, a BS with $256$ collocated antennas is placed at the center. To implement HmMIMO, we allocate one-fourth or half of the total antennas to the CBS, i.e., $N_b=64$ and $N_b=128$, marked by $HmMIMO-1/4$ and $HmMIMO-1/2$, respectively, while the remaining $192$ or $128$ APs are distributed randomly within the coverage area. The users within a distance of \SI{200}{\meter} to the CBS are treated as NUs while the others are FUs. At each simulation epoch, the locations of APs and users randomly change, and a total of $10^6$ epochs are conducted.

Large-scale fading is computed by $\beta=10^\frac{\mathcal{L}+\mathcal{X}}{10}$, where the shadowing $\mathcal{X}\sim \mathcal{N}(0,\sigma_{sd}^2)$ with $\sigma_{sd}=8\mathrm{dB}$. The path loss $\mathcal{L}$ is calculated by the COST-Hata model \cite{Ref_ngo2017cellfree}, taking values $d_0=10\mathrm{m}$, $d_1=50\mathrm{m}$, $f_c=2\mathrm{GHz}$, $h_{AP}=12\mathrm{m}$, and $h_{UE}=1.7\mathrm{m}$. Per-antenna and UE power constraints are set to $p_d=200\mathrm{mW}$ and $p_u=200\mathrm{mW}$, respectively.  The white noise power density equals $-174\mathrm{dBm/Hz}$ with a noise figure of $9\mathrm{dB}$, and the signal bandwidth equals $5\mathrm{MHz}$. The full-power strategy is applied for the DL transmission of all approaches. For example, CFmMIMO controls power like $\eta_{m}=\left(\sum_{k=1}^{K} \alpha_{mk} \right)^{-1}$, $\forall m$ and $\eta_{m}=\left(\sum_{k\in \mathbb{K}_F} \alpha_{mk} \right)^{-1}$ for the activated APs $\mathbb{M}_F$ in our proposed approach. 

Initially, \figurename \ref{fig:result1} displays the cumulative distribution functions (CDFs) of the sum capacity achieved by four different approaches in the uplink. Among these, mMIMO demonstrates the weakest performance, as users distant from the centralized antenna array experience very low data rates, thus diminishing the system capacity.  In our implementation, each user in UCmMIMO selects five nearby APs. As anticipated, UCmMIMO shows inferior performance compared with CFmMIMO because only a subset of APs participate in communications. However, this strategy offers the benefit of reduced fronthaul signaling. Our proposed approach clearly outperforms the three benchmarks, regardless of whether installing a quarter or half of the antennas at the CBS

The user-experienced data rate, as defined by 3GPP, is anchored in the $5^{th}$ percentile point ($5\%$) of the CDF. This metric provides a meaningful measurement of perceived performance at the cell edge.  In traditional cellular networks employing mMIMO, there exists a substantial performance gap between users at the cell edge and those at the cell center.  As shown in \figurename \ref{fig:result2}, the user-experienced rate with mMIMO approaches zero. As anticipated, CFmMIMO and UCmMIMO deliver consistent service quality, yielding user-experienced rates of about $0.47\mathrm{bps/Hz}$ and $0.49\mathrm{bps/Hz}$, respectively. But this improvement in worst-case performance comes at the expense of sacrificing the high performance typically enjoyed by cell-center users, as indicated by the circle in this figure. When considering $95^{th}$ percentile point of the CDF to identify achievable high rates, mMIMO yields approximately $7.88\mathrm{bps/Hz}$, compared to $2.96\mathrm{bps/Hz}$ and $2.25\mathrm{bps/Hz}$ of CFmMIMO and UCmMIMO, respectively. The user-experienced rates for HmMIMO-1/2 and HmMIMO-1/4 are $0.06\mathrm{bps/Hz}$ and $0.11\mathrm{bps/Hz}$, respectively, while their high rates reach $6.89\mathrm{bps/Hz}$ and $5.91\mathrm{bps/Hz}$. As designed, HmMIMO strikes a good balance between offering high rates and maintaining uniform service.

In the downlink, our proposal exhibits comparable superiority, as illustrated in \figurename \ref{fig:result3} and \figurename \ref{diagram:result2}. The user-experienced rate for mMIMO is close to zero, in comparison with $1.07\mathrm{bps/Hz}$ and $0.92\mathrm{bps/Hz}$ of CFmMIMO and UCmMIMO. HmMIMO-1/4 and HmMIMO-1/2 achieve almost the same DL performance, with a user-experienced rate of $0.07\mathrm{bps/Hz}$. Considering the $95^{th}$ percentile point of the CDF, mMIMO reaches $7.0\mathrm{bps/Hz}$, compared with $3.83\mathrm{bps/Hz}$ and $2.45\mathrm{bps/Hz}$ of CFmMIMIO and UCmMIMIO, respectively. Thanks to the coherent gain by using DL pilots, HmMIMO-1/2 and HmMIMO-1/4 achieve a result of $9.65\mathrm{bps/Hz}$, even better than that of mMIMO.

Last but not least, it is worth emphasizing that the performance superiority of our proposal doesn't necessarily come with increased complexity. As observed, its communication protocol is simple since the adaptation relies on large-scale fading, which varies slowly and is frequency-independent. This heterogeneous design offers additional technical merits, such as decreased power consumption and minimized fronthaul overhead, through the deactivation of distant APs.

\section{Conclusions}
This paper introduced a heterogeneous system design for cell-free massive MIMO, seamlessly integrating both co-located and distributed antennas. A central BS serves users in close proximity, providing them with high-rate connectivity. Distributed APs aid users far away from the BS, aiming to improve the worst-case service quality. By selecting the strongest AP for each far user, downlink pilots are enabled as the dimension of MIMO substantially degrades, thereby enhancing performance through coherently detecting signals. Numerical evaluations confirm that the heterogeneous design outperforms its homogeneous counterparts, exhibiting superior sum capacity while maintaining acceptable user-experienced rates. Moreover, the costs associated with AP sites and fiber cables are reduced. The power consumption of APs and the fronthaul signaling overhead are also lowered.

\bibliographystyle{IEEEtran}
\bibliography{IEEEabrv,Ref_COML}

\end{document}